\documentclass[%
 reprint,
superscriptaddress,
showpacs,
 amsmath,amssymb,
 aps,
]{revtex4-1}
\usepackage{graphicx}
\usepackage{color}
\usepackage{verbatim}

\begin{document}

\preprint{APS/123-QED}

\title{Quantum light propagation in longitudinally inhomogeneous media \\ as a  spatial Lewis-Ermakov physical invariance}

\author{David Barral}\email{Corresponding author. Email: david.barral@usc.es}
\author{Jes\'us Li$\tilde{\rm{n}}$ares}
\affiliation{Optics Area, Department of Applied Physics, Faculty of Physics and Faculty of Optics and Optometry, University of Santiago de Compostela, Campus Vida s/n (Campus Universitario Sur), E-15782 Santiago de Compostela, Galicia, Spain.}

\begin{abstract} We study the propagation of quantum states of light in separable longitudinally inhomogeneous media. By means of the usual quantization approach this kind of media would lead to the unphysical result of quantum noise squeezing. This problem is solved by means of generalized canonical transformations in a comoving frame. Under these transformations the generator of propagation is a physical Lewis-Ermakov invariant in space which is quantized and, accordingly, a propagator consistent with experiments is obtained. Finally, we show that the net effect  produced by propagation in these media is a quantum Gouy's phase with application in quantum interferometry.
\end{abstract}


\maketitle 

\section{Introduction}
 
In the last years great attention has been drawn to the time dependent quantum harmonic oscillator. This problem has been thoroughly studied both from mechanical \cite{Yeon1994, Guasti2003} and electromagnetic \cite{Pedrosa2009, Choi2004} points of view, showing squeezing effects. The approach most used in the tackle of this sort of problems is the use of quantum mechanical invariant operators, in particular the Lewis-Ermakov invariant \cite{Lewis1967}. 

In the field of guided propagation of light, the analogous problem is the study of longitudinally inhomogeneous (${\it LI}$) media. An invariant approach was extensively used in \cite{Krivoshlykov1992} and more recently in \cite{MoyaCessa2009} in the study of graded index media but, in both cases, they studied classical problems by means of a quantum-theoretical approach. Quantum propagation problems have been dealt with in homogeneous media, as shown in \cite{Luks2002} and references therein. These studies provided the background of the quantum theory of light propagation showing that the operator which describes correctly the quantum spatial propagation along an arbitrary direction $z$ is the Momentum operator $\hat{\mathcal{M}}$, since the Hamiltonian approach fails for problems like dispersive media, counterpropagation and ${\it LI}$ media as well \cite{Linares2008}. A first approximation to ${\it LI}$ media were carried out by Abram \cite{Abram1987} and Glauber and Lewenstein \cite{Glauber1991}, where single optical discontinuities between homogeneous media were analyzed. In these studies they showed the different physical behaviour the fields experience from the analogous mathematical problem of the time dependent quantum harmonic oscillator, proving that the field quadrature noise does not exhibit real squeezing, in agreement with experiments.

As far as we know, there is not any work dealing with quantum states of light propagating in ${\it LI}$ media in such a way that field quadratures do not exhibit real squeezing. So, our purpose is to give a phenomenological approach to this problem in waveguides. To this end, we will start obtaining the classical solutions which help us to find a reference frame which continuously performs a physical variable change in ${\it LI}$ media and eliminates the virtual squeezing, leading us to a proper propagation generator, a quantum Lewis-Ermakov type Momentum operator $\hat{\mathcal{M}}$. From this operator we will obtain Fock states in the optical-field strength (OFS) representation $\mathcal{E}$ and derive the propagator, proving the lack of squeezing in this representation and the arising of a quantum Gouy's phase. Furthermore, we will present the particular case of  propagation of a gaussian quantum state in a cosine-type ${\it LI}$ medium, where will be shown that the net effect of this kind of media on quantum states is the generation of a  quantum Guoy's phase dependent on features of both the media and the input quantum state.

\section{Classical analysis of propagation in longitudinally inhomogeneous media}

Our aim is to study the propagation of waveguided modes of quantum light in dispersion-free and non-magnetic media with separable inhomogeneous refractive index in an arbitrary direction of propagation $z$, given by \cite{Sodha1977}:
\begin{equation} \label{index}
n^{2}(x,y,z)= n_{0}^{2}\, f^{2}(x,y) + \Delta n^{2}\, h^{2}(z),
\end{equation}
where the longitudinal $h(z)$ and transversal $f(x,y)$ parts of the index are completely independent and $n_{0}$ and $\Delta n$ are constants. We focus on the separable index problem as it does not show coupling and therefore radiation modes (losses). 

From Maxwell equations, it is easy to show that the electric field $\boldsymbol{E}(x,y,z,t)$ obeys the vectorial wave equation \cite{Marcuse1974}:
\begin{equation}\label{VWE}
\mathbf{\nabla}^{2} \boldsymbol{E}_{t}+\mathbf{\nabla}_{t}( \boldsymbol{E}\,\mathbf{\nabla} (\ln n^{2})\,)=\frac{n^{2}}{c^{2}} \frac{\partial^{2} \boldsymbol{E}_{t}}{\partial t^{2}},
\end{equation}
with $\boldsymbol{E}=(E_{x}, E_{y}, E_{z})\equiv(\boldsymbol{E}_{t}, E_{z})$. Let us consider monochromatic guided $1D$ vector modes with frequency $\omega_{\sigma}$ represented by vector field solutions with the following factorable complex amplitudes:
\begin{equation} \label{ModesE}
\boldsymbol{E}_{t}(x,y,z,t)= \sum_{\sigma} q_{\sigma c}(z)\boldsymbol{\xi}_{t\sigma}(x,y)\,e^{-i\omega_{\sigma} t}, \\
\end{equation}
where we have used $\sigma$ for simplicity standing for the modal numbers $\nu$, $\mu$ in each transverse direction, z-dependent complex coefficients $q_{\sigma c}(z)$ fulfilling $\sum_{\sigma}\vert q_{\sigma c}(z) \vert^{2}=1$, and electric normalized transverse complex amplitudes $\boldsymbol{\xi}_{t\sigma} (x,y)$ corresponding to quasi-TE (or quasi-TM) modes fulfilling a {\it quasi}-complete orthonormalization condition \cite{Linares2008}, which belong to the homogeneous part of the refractive index and satisfy:
\begin{equation}\label{TransWaveEq}
\begin{split}
\nabla_{t}^{2} \, \boldsymbol{\xi}_{t\sigma} +& k_{0}^{2} n_{0}^{2} f^2 (x,y) \,\boldsymbol{\xi}_{t\sigma} +\\&\nabla_{t}( \boldsymbol{\xi}_{t\sigma}\,\nabla_{t} (\ln n_{0}^{2} f^2 (x,y)) = \beta_{t \sigma}^{2}\,\boldsymbol{\xi}_{t\sigma},
\end{split}
\end{equation}
with $ \beta_{t}$ the transverse propagation constant. Solutions of this equation give us the invariant transverse modal structure of the field. Applying equations \textcolor{black}{(\ref{index})}, (\ref{ModesE}) and (\ref{TransWaveEq}) into (\ref{VWE}), we obtain:
\begin{equation}\label{WaveEq}
\frac{d^{2}q_{\sigma}}{dz^{2}}+\beta_{\sigma}^{2}(z)\,q_{\sigma}=0,
\end{equation}
\textcolor{black}{where ${q}_{\sigma}=({q}_{\sigma c}+{q}_{\sigma c}^{*})/2$ stands for the real electric field coefficients}, $\beta^{2}_{\sigma}(z) = \beta_{t \sigma}^{2} + k_{0}^{2} \Delta n^{2} h^{2}(z)$ is the local propagation constant of the $\sigma$-mode \textcolor{black}{and where we have used the approximation $E_{z}\ll E_{x}, E_{y}$. It is important to outline that $E_{z}=0$ in the case of TE modes, that is, equation (\ref{WaveEq}) is exact for such modes.} This propagation equation clearly suggests a local spatial harmonic oscillator and therefore it can be directly derived from spatial-type Hamilton equations where the Hamilonian is substituted by the Momentum, since it is the generator of spatial translations \textcolor{black}{\cite{Abram1987,Linares2012},} given by:
\begin{equation} \label{MomentoOH}
 \mathcal{M}_{\sigma}= \frac{1}{2} [p_{\sigma}^{2} + \beta_{\sigma}^{2}(z)\,q_{\sigma}^{2}],
\end{equation}
with $p_{\sigma}=q_{\sigma}'$ and prime stands for $z$-derivative. This result is analogous to that obtained in \cite{Pedrosa2009} where time-dependent linear media was studied. The classical Momentum (\ref{MomentoOH}) is equivalent to the Hamiltonian of a time-dependent harmonic oscillator, with $\beta (z)$ playing the role of $\omega(t)$ \cite{Lewis1967}. 
Likewise, the solution of equation (\ref{WaveEq}) is easily obtained via the use of the complex electric field ${q}_{\sigma c}$ in the following way:
\begin{equation} \label{qc}
{q}_{\sigma c}(z)=\rho_{\sigma} \,e^{i\theta_{\sigma}} {q}_{\sigma c}(0),
\end{equation}
where $\rho_{\sigma}$ and $\theta_{\sigma}$ are real functions obtained by solving:
\begin{gather} \label{EL}
\frac{d^{2}\rho_{\sigma}}{dz^{2}}+\beta_{\sigma}^{2}(z)\rho_{\sigma}=\frac{\beta_{0 \sigma}}{\rho_{\sigma}^{3}}, \\  \label{ELtheta}
\frac{d\theta_{\sigma}}{dz}=\frac{\beta_{0 \sigma}}{\rho_{\sigma}^{2}},
\end{gather}
 with $\beta_{o \sigma} \equiv \beta_{\sigma} (0)$. Equation (\ref{EL}) is an Ermakov-Pinney equation with solutions given by \cite{Pinney1950}:
\begin{gather}\label{rhoz}
\rho_{\sigma}(z)=[(\beta_{o \sigma}\,u_{\sigma}(z))^{2} + v_{\sigma}^{2}(z)]^{1/2},\\ \label{rho0}
\rho_{\sigma}(0)=1, \quad \rho_{\sigma}'(0)=0,
\end{gather}
and where $u_{\sigma}$ and $v_{\sigma}$ are linearly independent functions that satisfy equation (\ref{WaveEq}) and have  the following initial conditions and Wronskian:
\begin{gather} \label{cond1}
{u_{\sigma}}(0)={v_{\sigma}'}(0)=0, \\  \label{cond2}
{u_{\sigma}'}(0)={v_{\sigma}}(0)=1, \\  \label{Wronskian}
{{W_{\sigma}}}={u_{\sigma}'}\,{v_{\sigma}}-{v_{\sigma}'}{u_{\sigma}}=1.
\end{gather}
So, the classical wave amplitude changes continously during propagation by a factor $\rho_{\sigma}$, whereas the modal propagation constant is given by $\tilde{\beta_{\sigma}}\equiv\theta'_{\sigma}$ via equation (\ref{ELtheta}) where $\theta_{\sigma}$ represents the total phase accumulated in the propagation.  

\section{Quantization in longitudinally inhomogeneous media}

The next step is to quantize the classical Momentum (\ref{MomentoOH}). But before carrying out this step, some considerations have to be taken into account. Direct quantization of the classsical fields $q_{\sigma}$ and $p_{\sigma}$ would lead to a quantized $z$-dependent Momentum analogous to the Hamiltonian for a time-dependent harmonic oscillator and, therefore, following the usual steps, we would have propagation equations leading to quadrature noise squeezing \cite{Yeon1994, Guasti2003, Pedrosa2009, Choi2004}. But in the sudy of propagation in $\it{LI}$ dielectric media, this approach does not provide consistent results. As was pointed out by Abram in his seminal paper about quantization of light in dielectric media  \cite{Abram1987}, when quantum states propagating in a dielectric are represented in the basis of free-space photons, they seem to be squeezed, but inside a dielectric there is no experiment that can detect free-space photons. This happens because if we wish to detect photons inside the medium the fields would experiment an effective refractive index equivalent to the squeezing parameter and, therefore, a scale change is necessary. In the same spirit an interesting discussion about quantization in a dielectric was carried out by Glauber and Lewestein in \cite{Glauber1991}. In this study is stressed that the measurement of quantum states of light is based on photoabsorption processes carried out in the basis which diagonalizes the Hamiltonian in every medium. This is so because the photocount distribution is given by a normal ordered correlation product in the local basis, result of the physical property of the photoabsorption process that the energy of the field decreases when a photon is absorbed. But in any other basis this is not true, because of the mixing of absorption and emission operators, and therefore of frequencies, lacking the normal ordering in the correlation product. Thereby the state in these bases can not be measured and the photons are so-called virtual and, accordingly, the squeezing is virtual as well. This statement is closely related with Abram's idea and has to be applied to $\it{LI}$ media as it is the continuous limit of single discontinuities.

Hence, we look for some transformation which takes the classical Momentum (\ref{MomentoOH}) to be a constant of motion suitable to be quantized and simultaneously get rid of the virtual squeezing in a continuous way. Since carrying out the usual approach \cite{Yeon1994, Guasti2003, Pedrosa2009, Choi2004} squeezing proportional to the function $\rho(z)$ would appear, following the idea of a scale change in the discontinuity \cite{Abram1987, Glauber1991} we propose the following generalized canonical transformations from a generating function $G_{2}(q,\tilde{P})= \tilde{P}q/\rho$ and $z$ transformation \cite{Chetouani1989}:
\begin{equation}\label{GCT}
\tilde{Q}_{\sigma}=\frac{q_{\sigma}}{\rho_{\sigma}},    \quad   \tilde{P}_{\sigma}= \rho_{\sigma} p_{\sigma}, \quad s_{\sigma}=(\int^{z}_{0} \rho_{\sigma}^{2}(\textsl{z})\, d\textsl{z})^{-1};
\end{equation}
and the next gauge transformations related to the new $s$-frame:
\begin{equation}\label{GCT2}
Q_{\sigma}=\tilde{Q}_{\sigma},    \qquad  P_{\sigma}= \tilde{P}_{\sigma} - \frac{\dot{\rho}_{\sigma}}{\rho_{\sigma}} \tilde{Q}_{\sigma},
\end{equation}
with the dot standing for an $s$-derivative. Applying these relations into (\ref{MomentoOH}), we directly obtain the following Momentum:
\begin{equation} \label{MomentoGCT}
\mathcal{M}_{\sigma}(Q,P,s)= \frac{1}{2} [P_{\sigma}^{2}(s) + \beta_{0\sigma}^{2}\,Q_{\sigma}^{2}(s)],
\end{equation}
where we have used the auxiliar equation (\ref{EL}). It is remarkable that these new phase space and longitudinal coordinates ($Q, P, s$) are analogous to the comoving coordinates and the conformal (or arc-parameter) time used in cosmology and, likewise, the canonical momentum transformation (\ref{GCT2}) is related to the Hubble's law \cite{Misner1973}. After these transformations, the continuous change in the value of the harmonic potential appears as constant in the new comoving frame of reference \cite{Takagi1990}. So, in this frame, there are not scale changes differentially and the medium seems to be homogeneous. The generalized canonical transformations eliminate the squeezing at each plane $z$ and none signature of it is obtained. These insights lead us to consider the comoving frame as the physical one and therefore the proper one for quantization. Furthermore, interestingly, if we rewrite this Momentum in the original phase space, we obtain:
\begin{equation} \label{MomentoOHEL}
\mathcal{M}_{\sigma}(q,p,z)= \frac{1}{2} [(\rho_{\sigma}\, p_{\sigma} - q_{\sigma}\, \rho_{\sigma}')^{2}+\beta_{0\sigma}^{2}(q_{\sigma}/\rho_{\sigma})^{2}].
\end{equation}
This is the space-analog of an Ermakov-Lewis invariant \cite{Lewis1967}. So, it can be said that the Momentum (\ref{MomentoGCT}) is a spatial Ermakov-Lewis invariant in a comoving frame and that this invariant is the generator of spatial propagation in this kind of media.

Now, following the principle of quantization of quantum mechanics $(Q_{\sigma}, P_{\sigma})\rightarrow (\hat{Q}_{\sigma}, \hat{P}_{\sigma})$, where $[\hat{Q}_{\sigma},\hat{P}_{\sigma'}]=i\hbar\delta_{\sigma, \sigma'}$, we obtain the Momentum operator:
\begin{equation} \label{QMomento}
\hat{\mathcal{M}}_{\sigma}= \frac{1}{2} [\hat{P}_{\sigma}^{2}(s) + \beta_{0\sigma}^{2}\,\hat{Q}_{\sigma}^{2}(s)].
\end{equation}
In order to obtain the eigenvalues and eigenfunctions of this operator, we define the following annihilation and creation operators:
\begin{gather}\label{dest}
\hat{A}_{\sigma}(s)=\frac{1}{\sqrt{2\hbar\beta_{0\sigma}}}[\beta_{0\sigma}\hat{Q}_{\sigma}+i\hat{P}_{\sigma}],  \\  \label{creat}
\hat{A}^{\dag}_{\sigma}(s)=\frac{1}{\sqrt{2\hbar\beta_{0\sigma}}}[\beta_{0\sigma}\hat{Q}_{\sigma}-i\hat{P}_{\sigma}]. 
\end{gather}
Therefore, equation (\ref{QMomento}) in terms of (\ref{creat}) can be rewritten as:
\begin{equation} \label{QMomento2}
\hat{\mathcal{M}}_{\sigma}= \hbar \beta_{0\sigma}[\hat{A}^{\dag}_{\sigma}\hat{A}_{\sigma}+1/2].
\end{equation}
The eigenstates of this Momentum are the same as those of the Number operator $\hat{N}_{\sigma}=\hat{A}^{\dag}_{\sigma}\hat{A}_{\sigma}$, which fulfills the next eigenvalue equation:
\begin{equation}
\hat{N}_{\sigma} \vert N_{\sigma} \rangle = N_{\sigma} \vert N_{\sigma} \rangle,
\end{equation}
where $N_{\sigma}$ and $\vert N_{\sigma} \rangle$ are eigenvalues and eigenstates for the Number operator, respectively. In terms of quantum-optical units \cite{Loudon1982}, we can redefine the quantum comoving scaled position $\hat{Q}_{\sigma}$ as the OFS operator $\hat{\mathcal{E}}_{\sigma}$, whose eigenstates are those measured in homodyne experiments \cite{Vogel1990, Schleich2001} and central to generalized quantum polarization \cite{Linares2011, Barral2013}, that is $\hat{\mathcal{E}}_{\sigma}=\sqrt{\beta_{0 \sigma }/2\hbar}\, \hat{Q}_{\sigma}$. The eigenstates $\vert N_{\sigma} \rangle$ in the optical field-strength $\mathcal{E}$ basis are given in terms of Hermite-Gauss functions as:
\begin{equation}
\Psi_{N}(\mathcal{E}_{\sigma},0)= \frac{2^{1/4}}{\sqrt{2^{N}\pi^{1/2}\,N!}}\,H_{N}(\sqrt{2}\mathcal{E}_{\sigma})\,e^{-\mathcal{E}^{2}_{\sigma}},
\end{equation}
and their quantum propagation is given by a spatial Schr\"odinger equation:
\begin{equation}\label{Schrodinger}
\hat{\mathcal{M}}_{\sigma} \,\Psi_{N} \equiv \hbar\beta_{0\sigma}\,[-\frac{1}{4}\frac{\partial^{2}}{\partial{\mathcal{E}_{\sigma}}^{2}}+\mathcal{E}^{2}_{\sigma}] \, \Psi_{N} = \hbar\beta_{0\sigma}\,\Psi_{N}.
\end{equation}
The solution of this equation yields the following propagated eigenstates $\vert N_{\sigma} \rangle$:
\begin{equation}\label{Number}
\Psi_{N}(\mathcal{E}_{\sigma}; s)=\frac{2^{1/4} \,e^{ i(N+1/2)\theta_{\sigma}}}{\sqrt{2^{N}\pi^{1/2}\,N!}}\,H_{N}(\sqrt{2}\mathcal{E}_{\sigma})\,e^{-\mathcal{E}^{2}_{\sigma}  } ,
\end{equation}
with $\theta_{\sigma}=\beta_{0\sigma} s_{\sigma}$ the total accumulated phase, in the same way as a quantum state propagating in a homogeneous medium, but in this case in a comoving frame, which is the physical one for this kind of problems as shown above. Because of the similarity of this phase with the classical Gouy's phase \cite{Boyd}, we call it spatial quantum Gouy's phase.

\section{Propagation of quantum light in longitudinally inhomogeneous media}

Now, in order to gain insight into the behaviour of the longitudinally inhomogeneous waveguide, we study the spatial propagation of gaussian states, like the coherent and squeezed quantum states of light, in the OFS representation. From the wavefunction (\ref{Number}) and Mehler's formula \cite{Yeon1994}, we easily obtain the propagator of the system for every mode \cite{Linares2012}:
\begin{equation*}
K ( \boldsymbol{\mathcal{E}}, \boldsymbol{\mathcal{E}_{0}}; z, 0 )=(\frac{i}{\pi})^{N/2} \prod_{\sigma}\, (\sin\theta_{\sigma})^{-1/2} 
\end{equation*}
\vspace{-0.2cm}
\begin{equation} \label{Propagator2}
e^{-\frac{i}{\sin\theta_{\sigma}} [\cos \theta_{\sigma}\,(\mathcal{E}_{\sigma}^{2} + \mathcal{E}_{0 \sigma}^{2}) - 2 \mathcal{E}_{\sigma} \mathcal{E}_{0 \sigma}]    },
\end{equation}
and the wavefunction of any pure state at the end of the medium can be worked out by means of:
\begin{equation} \label{PropWaveFunc}
\Psi(\boldsymbol{\mathcal{E}}; z)= \int K ( \boldsymbol{\mathcal{E}}, \boldsymbol{\mathcal{E}_{0}}; z, 0 )\, \Psi(\boldsymbol{\mathcal{E}_{0}}; 0)\, d\boldsymbol{\mathcal{E}_{0}},
\end{equation} 
where $\boldsymbol{\mathcal{E}_{0}}$ and $\boldsymbol{\mathcal{E}}$ stand for the $N$-mode field-strength at the beginning and any point of the $z$-dependent medium, respectively.

\vspace{-0.5cm}
\begin{figure}[h]
\centering
\includegraphics[width=0.51\textwidth]{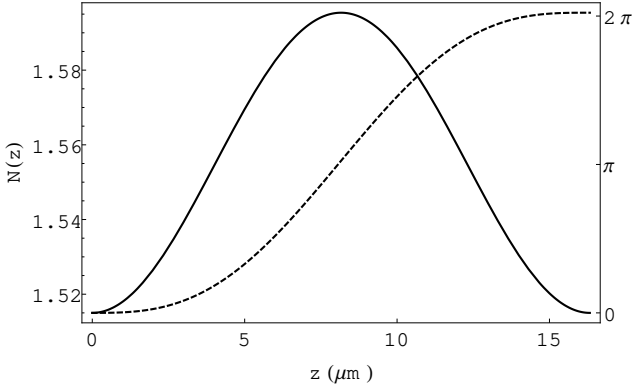}
\vspace {0cm}\,
\hspace{0cm}\caption{\label{F1}\small{Cosine-type effective refractive index $N(z)\equiv\beta(z)/k_{0}$ (solid line, left) and total accumulated classical phase $\theta$ (dash line, right) versus propagation distance $z$, for parameters $N_{t}\equiv\beta_{t}/k_{0}=1.515$, $\Delta n=0.5$ and $\Lambda=k_{0}/50$ $nm^{-1}$ for a wavelength $\lambda=653$ nm.}}
\end {figure}

A single-mode gaussian state $\vert \alpha \rangle$, where $\alpha= \vert \alpha \vert e^{i\phi}$, in the OFS space is given by \cite{Linares2012}:
\begin{equation} \label{CoherentIn}
\Psi(\mathcal{E}_{0}; 0)=(\frac{2}{\pi\,\Delta\mathcal{E}_{0}^{2}})^{1/4} exp\{\frac{-[\mathcal{E}_{0}-\vert \alpha \vert \,\cos\phi ]^2}{\Delta\mathcal{E}_{0}^{2}} \}\,e^{-i\delta_{0}},
\end{equation}
where $\delta_{0}=\sin \phi\, [\vert \alpha \vert^{2} \cos \phi - 2\vert \alpha \vert \mathcal{E}_{0}]$, $\phi=-\omega t$ is the phase associated to the temporal evolution of the state at every plane $z$ and $\Delta\mathcal{E}^{2}_{0}$ is the quantum noise or squeezing factor, with a value of one for a coherent state or different from one for squeezed states. Inserting equations (\ref{Propagator2}) and (\ref{CoherentIn}) in (\ref{PropWaveFunc}), after a long but straightforward calculation, we obtain the wavefunction at any point $z$ of the waveguide:
\begin{equation}\label{SqueezedOutFinal}
\begin{split}
\Psi(\mathcal{E}; z)&=(\frac{2}{\pi\,\Delta\mathcal{E}^{2}})^{1/4} \,e^{ i\Theta/2} e^{-i\delta} \\
&exp\{ \frac{-1}{\Delta\mathcal{E}^{2}}  [\mathcal{E}-\vert \alpha \vert\, \cos(\phi+\theta)\, ]^2 \},
\end{split}
\end{equation}
where we have denoted $\theta_{\sigma}\equiv\theta$ and defined $\delta=\sin (\phi+\theta)\, [\vert \alpha \vert^{2} \cos (\phi+\theta) - 2\vert \alpha \vert \mathcal{E}]$ and:
\begin{gather}\label{sq}
\Delta\mathcal{E}=[\frac{\sin^{2}\theta}{\Delta\mathcal{E}_{0}^{2}} + \cos^{2}\theta\,\Delta\mathcal{E}_{0}^{2}]^{1/2}, \\ \label{phasesq}
\Theta=arctan[\frac{\tan\theta}{\Delta\mathcal{E}_{0}^{2}}].
\end{gather}
Therefore, a coherent state keeps its quantum noise constant through propagation, $\Delta\mathcal{E}^{2}=\Delta\mathcal{E}_{0}^{2}=1$, and gets a Gouy's phase equal to the classical accumulated phase given by (\ref{ELtheta}), $\Theta=\theta$, as expected, unlike what we would obtain via the usual approach. On the other hand, an input squeezed state is both amplitude and phase modulated by the inhomogeneous medium, leading to different squeezing oscillations from those corresponding to an homogeneous medium, and with a squeezing factor given by ($\ref{sq}$). It is important to outline that the phase acquired by a squeezed state ($\ref{phasesq}$) is not predicted by the classical theory.

\begin{figure}[h]
\centering
\includegraphics[width=0.46\textwidth]{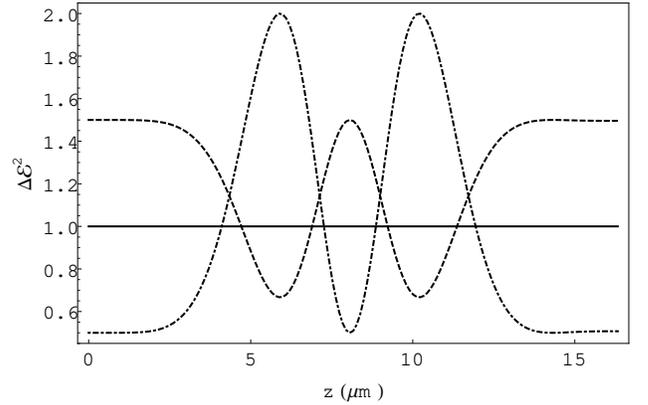}
\vspace {0cm}\,
\hspace{0cm}\caption{\label{F2}\small{Evolution of the quantum noise $\Delta \mathcal{E}^{2}$ in a cosine-type longitudinally inhomogeneous medium for coherent (solid line) and squeezed (dash and dash-dot lines) states of light.}}
\end {figure}

Next, we show a specific example of the effects this kind of medium produce over a single-mode gaussian quantum state like the above introduced. We present a longitudinally inhomogeneous cosine-type medium with $h(z)=\cos(\Lambda z)$ and length $L=2\pi/\Lambda$, as that depicted in Figure 1. The solution of equation (\ref{EL}) is given in terms of linearly independent Mathieu functions $u(z)$ and $v(z)$ \cite{Casperson1985}. Applying these solutions into equations (\ref{ELtheta}) and (\ref{rhoz}), we can obtain numerically the value of $\theta$ and, therefore, the squeezing factor (\ref{sq}) and the Gouy's phase (\ref{phasesq}), which characterize the output quantum state after propagation in this medium. In Figures 2 and 3 we show the behaviour of the Gouy's phase $\Theta$ and the quantum noise $\Delta\mathcal{E}^{2}$, respectively, for three different input quantum states: the solid line stands for the propagation of a coherent state where $\Delta\mathcal{E}_{0}^{2}=1$ and the dash and dash-dot lines for squeezed states with $\Delta\mathcal{E}_{0}^{2}=3/2$ and $\Delta\mathcal{E}_{0}^{2}=1/2$, respectively. As can be checked in Figure 2, the planes where maximum or minimum squeezing of the quantum noise are produced depend on the accumulated phase $\theta$, being different from those corresponding to an homogeneous medium, and the value of the input quantum noise $\Delta\mathcal{E}_{0}^{2}$. Likewise, as was outlined above, the classical equations do not predict the behaviour of states far from the classical, since squeezed and coherent states get different amounts of Guoy's phase $\Theta$, as is sketched in Figure 3. The classical phase $\theta$ is recovered at the planes of maximum and minimum squeezing. {Finally, we stress the interest this phase may have in quantum interferometry, since with an appropriate $\it{LI}$ medium in one arm of an interferometer, quantum interference can be adjusted controlling the strength of the quantum noise in a squeezed input state via equation (\ref{phasesq}).}

\begin{figure}[h]
\centering
\includegraphics[width=0.46\textwidth]{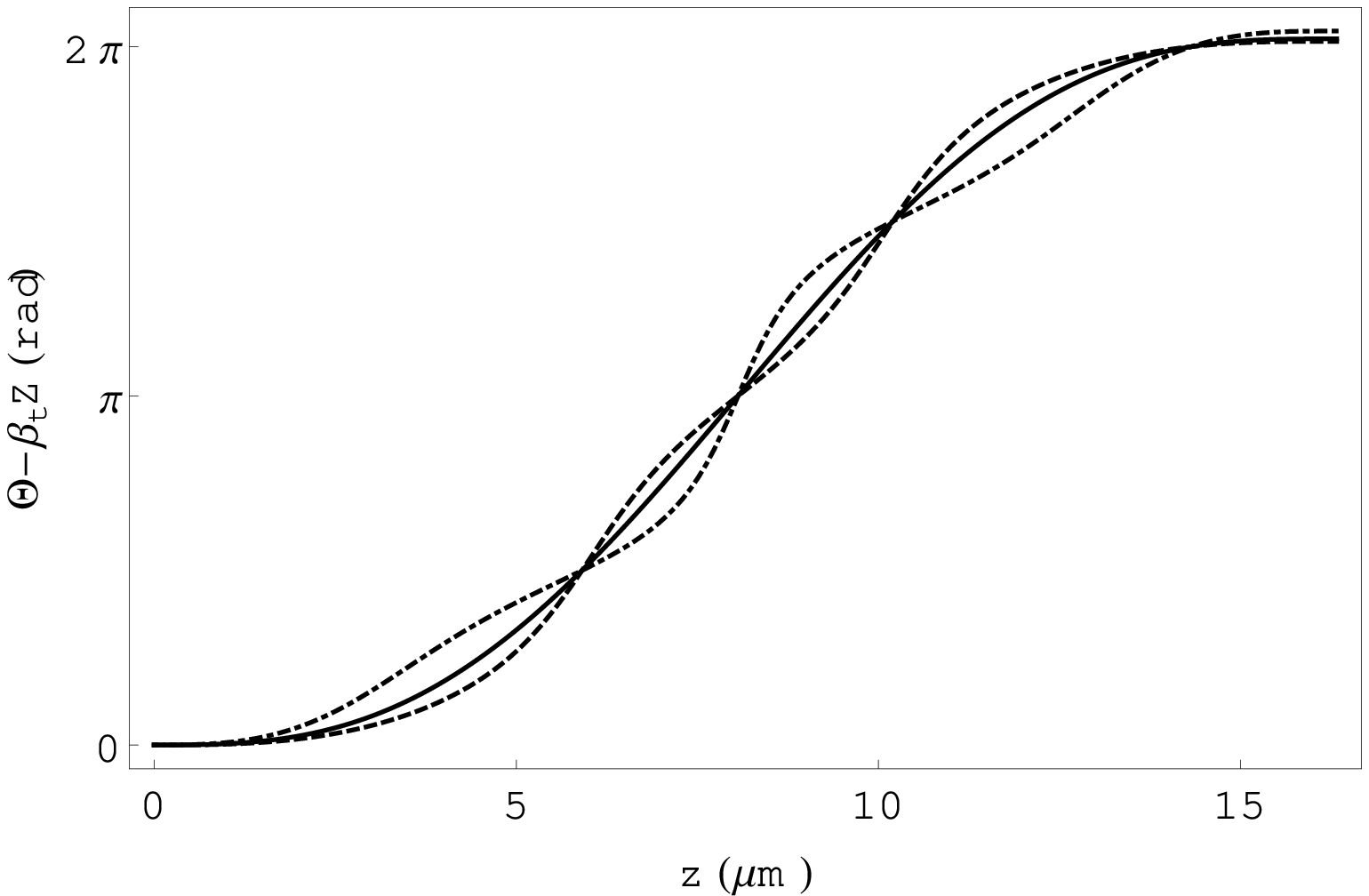}
\vspace {0cm}\,
\hspace{0cm}\caption{\label{F3}\small{Evolution of the quantum Gouy's phase $\Theta$ in a cosine-type longitudinally inhomogeneous medium for coherent (solid line) and squeezed (dash and dash-dot lines) states of light.}}
\end {figure}

\section{Summary}
We have studied the propagation of quantum states of light in separable longitudinally inhomogeneous media. We have proposed canonical transformations in a comoving frame and derived the generator of propagation, the classical Momentum, realizing that it is a spatial Lewis-Ermakov invariant. We have quantized it and obtained the propagator in the OFS representation which is physically consistent. Furthermore, we have presented the propagation of a gaussian quantum state of light in a cosine-type refractive index medium and shown the net effect produced by this family of media: a quantum Gouy's phase dependent on the input quantum state with application in quantum interferometry.

\end{document}